\documentclass[runningheads]{llncs}
\usepackage{amsmath,amsfonts}
\usepackage{verbatim}
\usepackage[colorlinks,
            linkcolor=red,
            citecolor=green
            ]{hyperref}
\usepackage[misc]{ifsym}
\usepackage{graphicx}
\graphicspath{{figure2/}}

\begin{document}
\title{VHS to HDTV Video Translation using Multi-task Adversarial Learning}%
%
\author{Hongming Luo\inst{1,2,3,4}  \and
Guangsen Liao\inst{1,2,3,4} \and Xianxu Hou\inst{1,2,3,4} \and Bozhi Liu\inst{1,2,3,4} \and Fei Zhou \inst{1,2,3,4}(\Letter) \and Guoping Qiu\inst{1,2,3,4,5}}

\authorrunning{H. Luo et al.}

%
\institute{ College of Electronics and Information Engineering, Shenzhen University, Shenzhen, China \and 
Guangdong Key Laboratory of Intelligent Information Processing, Shenzhen, China \and Guangdong Laboratory of Artificial Intelligence and Digital Economy (SZ), Shenzhen, China \and Shenzhen Institute of Artificial Intelligence and Robotics for Society, Shenzhen, China 
\and School of Computer Science, University of Nottingham, Nottingham, UK
\\ \email {flying.zhou@163.com} }
\maketitle              
\begin{abstract}

There are large amount of valuable video archives in Video Home System (VHS) format. However, due to the analog nature, their quality is often poor. Compared to High-definition television (HDTV), VHS video not only has a dull color appearance but also has a lower resolution and often appears blurry. In this paper, we focus on the problem of translating VHS video to HDTV video and have developed a solution based on a novel unsupervised multi-task adversarial learning model. Inspired by the success of generative adversarial network (GAN) and CycleGAN, we employ cycle consistency loss, adversarial loss and perceptual loss together to learn a translation model. An important innovation of our work is the incorporation of super-resolution model and color transfer model that can solve unsupervised multi-task problem. To our knowledge, this is the first work that dedicated to the study of the relation between VHS and HDTV and the first computational solution to translate VHS to HDTV. We present experimental results to demonstrate the effectiveness of our solution qualitatively and quantitatively. 
\end{abstract}

\section{Introduction}
\label{sec:sec1}
  With the rapid development of electronic technology, especially video display technology, the resolution of television and display is higher and higher as well as the expanding of color space. Nowadays, high-definition television is widespread and 4K television, even HDR television is gradually available. Compared with the development of video display technology, video resources that can suitable for such high-resolution display are very scarce. Massive video resources have been generated in the long-term accumulation. Some classic old movies, programmes as well as some important events and moments are recorded through a large amount of video or image data, but these precious old resources can't reach the level of 4K or HDR display. If these resources are not processed, playing on the above display device will result in very poor visual experience.
  
\begin{figure}[!t]

\begin{minipage}{0.42\linewidth}
\centerline{\includegraphics[height=3.2cm,width=5cm]{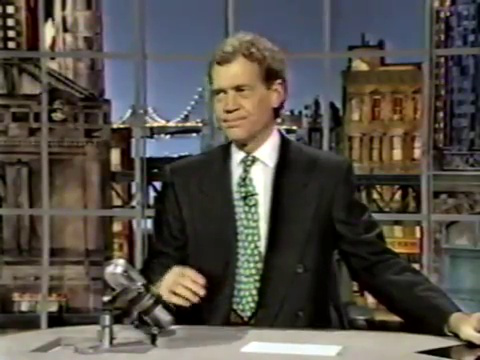}}
\centerline{(a) VHS frame}
\end{minipage}
\hfill
\begin{minipage}{0.6\linewidth}
  \centerline{\includegraphics[height=3.2cm,width=6.5cm]{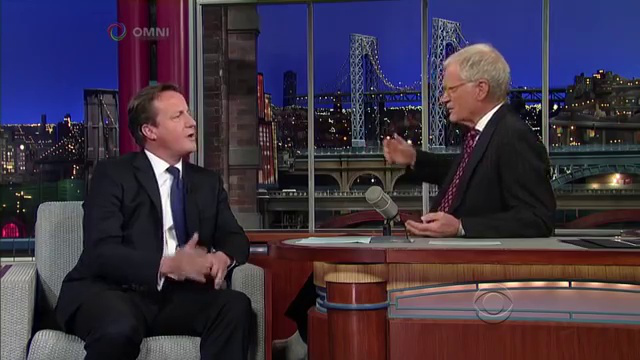}}
  \centerline{(b) HDTV frame}
\end{minipage}

\caption{Illustration of VHS and HDTV frames.}
\label{fig:res1}
\end{figure}

   In addition, the original imaging equipment and storage methods not only determine the low resolution of these video resources, but also their poor color performance. For example, Video Home System (VHS) was the dominant video format widely used across the world since 1970s before it was replaced by high-definition television (HDTV) in the new millennium. Due to the analog nature of VHS recording medium and limited storage memory, the visual quality of VHS video is inferior. As shown in Fig.~\ref{fig:res1}(a), VHS frame has low resolution and low color contrast. Hence, it is very important to effectively process these resources to give users a better visual experience on existing high-definition television. 
   
   The most perfect video enhancement and restoration technology is transforming shaky, blurry, color-distorted, noisy, overly dark or bright and low resolution video footage into sharper, clearer and visually pleasing videos. However, most of existing technology only focus on single problem mentioned above and there are corresponding training pairs, which is to say that there are ground truth. But in reality, we sometimes can't reach that situation. To translate the video frames like Fig.~\ref{fig:res1}(a) to the video frames like Fig.~\ref{fig:res1}(b), we need video frames with high resolution and high color contrast corresponding to VHS video frames. Although we can get HDTV video frames, they are not pixel-wise corresponding to VHS video frames which means we lack of ground truth for training(More details are introduced in section~\ref{sec:sec3}). Based on this problem, we propose a method that settle unsupervised multi-task problem. Our goal is learning a mapping from the video frames like Fig.~\ref{fig:res1}(a) to the video frames like Fig.~\ref{fig:res1}(b). Firstly, we use generative adversarial network to translate VHS video frames to HDTV video frames. At the same time, the cycle consistency loss is needed to keep the contents of video frames. Only HDTV video frames can be used for training super-resolution model due to the lack of VHS paired training samples from low resolution to high resolution. But the discrepancy between VHS video frames and HDTV video frames makes the model trained on HDTV video frames not applicable for VHS video frames. Therefore, we need a method to make the model have better applicability to both training samples. The most simple way to achieve that is sharing weights of models training for two kinds of samples, so as to achieve the effect of the model to complete the color conversion and super-resolution with single model.
Our contribution can be summarized as:
\begin{itemize}
\item As far as we known, this is the first work to focus on translating VHS video to HDTV video which is a unsupervised multi-task problem.
\item We propose a novel multi-task adversarial learning model for translating VHS to HDTV video and set up a framework that can solved unsupervised multi-task problem. Our experimental results to demonstrate our approach's effectiveness.
\end{itemize}

\section{Related Works}
\label{sec:sec2}
\subsubsection{Style Transfer} 
After Gatys et al.~\cite{gatys2016image} successfully apply convolutional neural networks(CNN) in style transfer, many work based on CNN emerge. Gatys et al.~\cite{gatys2016image} use pre-trained model to extract features of content and style, then train a generative network to synthesize stylized images through iterative optimization. The speed of this strategy is slow because one network is trained for one image. Subsequently, Johnson et al.~\cite{johnson2016perceptual} propose a feed-forward network to generate styled image by decreasing both style and content features losses, which increase the speed a lot. However, all these works are kept the features unchanged between generated image and content image. Even though the style image is natural image, the generated image has some non-photorealistic artifacts. Hence, there are some work focusing on photorealistic style transfer. Deep photo style transfer, proposed by Luan et al.~\cite{luan2017deep}, augments style loss with semantic segmentation and adds photorealism regularization to make generated image look more photorealistic. After that, photoWCT~\cite{li2018closed}, uses photorealistic variant of WCT to replace VGG network's upsampling module and applies additional post-processing to reduce artifacts. In addition, generative adversarial network(GAN) is used in CycleGAN~\cite{zhu2017unpaired} to transfer image style, and cycle consistency loss is used to constrain the content information.

\subsubsection{Super Resolution} 
Recently, deep neural networks have achieved great success in the field of super resolution. Deep network cascade for super resolution~\cite{cui2014deep} uses multi-level autoencoder to achieve super resolution. And Dong et al.~\cite{dong2015image} use convolutional neural network(CNN) simulate each module in traditional super resolution task with different convolutional layers: low-resolution feature extraction, low-resolution to high-resolution feature mapping and high-resolution image reconstruction. But when the network gets deeper, the performance is not further improved. In order to improve the efficiency of network, Dong et al.~\cite{dong2016accelerating} take low-resolution image instead of bilinearly interpolated image as input, and design a deconvolutional layer at the end of network to complete the upsampling of image features. To achieve deeper network, Kim et al.~\cite{kim2016accurate} propose VDSR(very deep super resolution) model with a skip-connection to transmit low frequent signal. DRRN~\cite{tai2017image}(deep recursive residual network) recursively calls the same residual module whose parameters are shared, so it makes the network deeper without increasing number of parameters. RDN~\cite{zhang2018residual}(residual dense network) utilizes both residual block and dense block, which is more complex skip-connection to achieve super resolution. From these cases, deeper networks, better performance. As the development of generative adversarial network, Ledig et al.~\cite{ledig2017photo} define adversarial loss, and acquire better perceptual results using adversarial training strategy. 

\section{Adversarial Learning for VHS to HDTV }
\label{sec:sec3}

\begin{figure}[htb]

\begin{minipage}[b]{1.0\linewidth}
  \centering
  \centerline{\includegraphics[width=12cm]{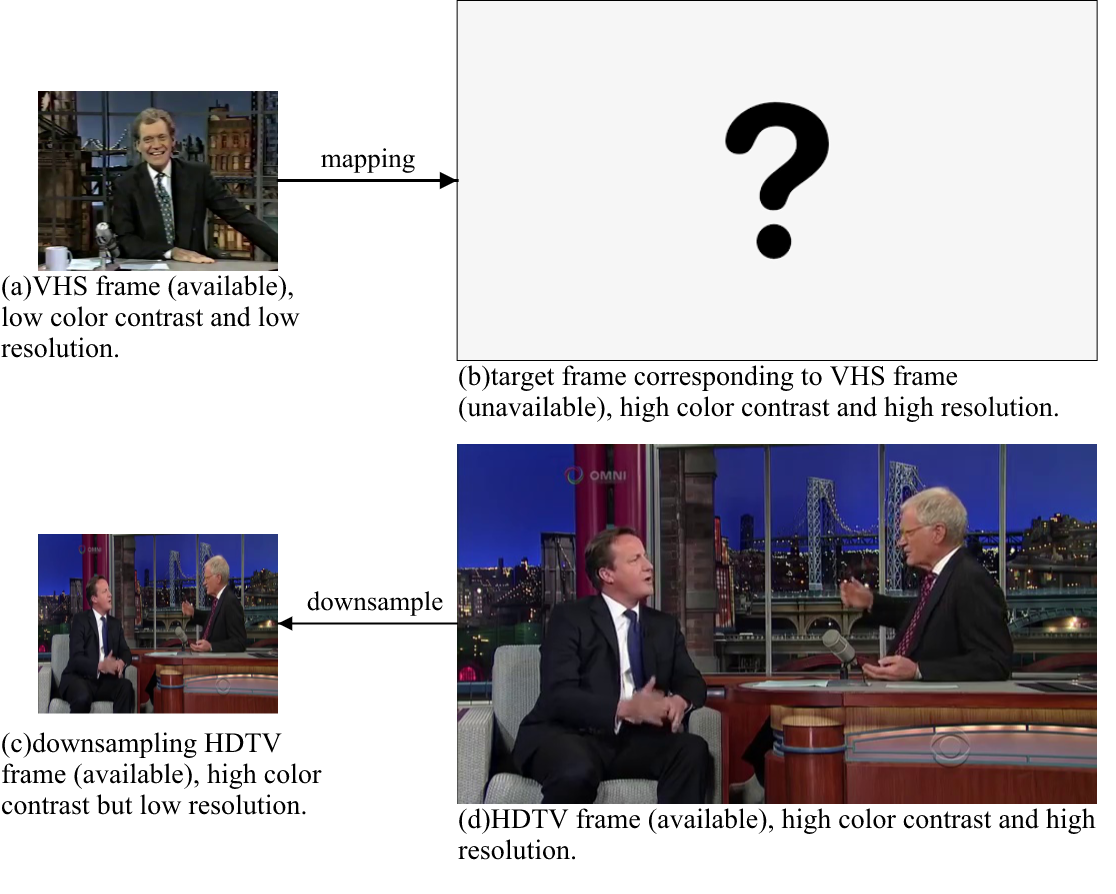}}
  
\end{minipage}
\caption{Problem description.}
\label{fig:res2}
\end{figure}

As shown in Fig.~\ref{fig:res2}, our goal is to learn a mapping function from the domain of VHS videos like Fig.~\ref{fig:res2}(a) with low color contrast and low resolution to the domain of target videos like Fig.~\ref{fig:res2}(b) with high color contrast and high resolution. And Fig.~\ref{fig:res2}(b) should be similar with HDTV videos like Fig.~\ref{fig:res2}(d). We can obtain high color contrast and high resolution HDTV frames like Fig.~\ref{fig:res2}(d) and downsample HDTV frames to low resolution frames like Fig.~\ref{fig:res2}(c). Using Fig.~\ref{fig:res2}(a), (c) and (d), we can learn a mapping from Fig.~\ref{fig:res2}(a) to Fig.~\ref{fig:res2}(c), which is only enhance color contrast, and a mapping from Fig.~\ref{fig:res2}(c) to Fig.~\ref{fig:res2}(d), which is only enhance resolution. However, the target frames like Fig.~\ref{fig:res2}(b) corresponding to VHS frames are unavailable, so that there is no way to learning a mapping from Fig.~\ref{fig:res2}(a) to Fig.~\ref{fig:res2}(b) directly. Hence, this problem is divided into an unsupervised multi-task problem category, which we need to address not only the color style but also the resolution of videos without ground truth. Our multi-task adversarial learning model for translating VHS video to HDTV video will be introduced below.

 We use $x$ and $y$ to represent the VHS(Fig.~\ref{fig:res2}(a)) and HDTV frames(Fig.~\ref{fig:res2}(d)) respectively, and $z$ is the low-resolution counterpart(Fig.~\ref{fig:res2}(c)) of $y$. The distributions of $x$ and $y$ are denoted as $x \sim p_{data}(x)$ and $y \sim p_{data}(y)$ respectively.

\subsection{Model Architecture}
\begin{figure}[htb]

\begin{minipage}[b]{1.0\linewidth}
  \centering
  \centerline{\includegraphics[width=10cm]{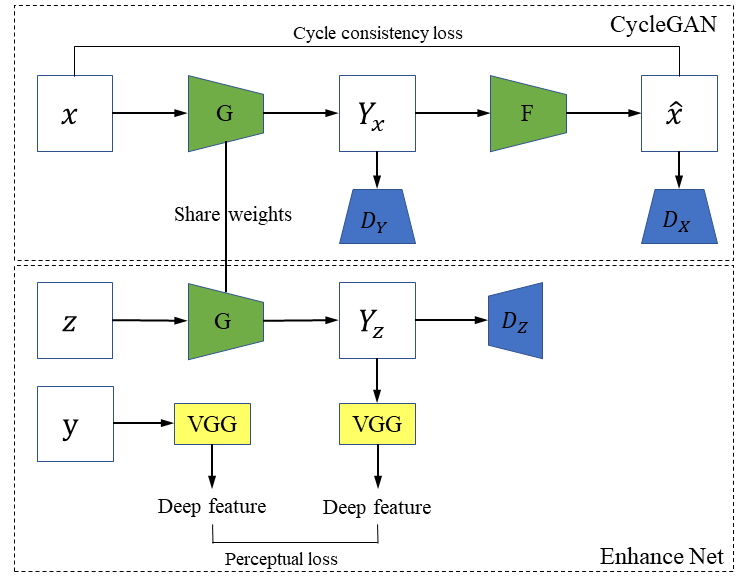}}
  
\end{minipage}
\caption{Architecture of our model.}
\label{fig:res3}
\end{figure}

The architecture of our model is given in Fig.~\ref{fig:res3}, which consists of two parts. Each part is denoted by a dashed box in Fig.~\ref{fig:res3}. The top part is the same as CycleGAN ~\cite{zhu2017unpaired}, where we define two generators $G:x\to Y_{x}$ and $F:Y_{x}\to x $ as two mappings as well as two discriminators $D_{Y}$ and $D_{X}$. $D_{Y}$ aims to distinguish images generated by $G$ and HDTV frames while $D_{X}$ aims to distinguish images generated by $F$ and VHS frames. The generator in the bottom part, i.e., the Enhance Net, shares weights with the generator $G$ above. And we add a perceptual loss between $y$ and generated $Y_{z}$, which is computed via features extracted by the pre-trained (and fixed) VGG19~\cite{simonyan2014very}. In addition, we also define another discriminator $D_{Z}$, which aims at distinguishing images generated by $G$ and HDTV frames. The difference of the generated images and the ground truth lies in spatial resolution only. $Y_{x}$ is image that contains the content of $x$ with $Y$ style while $Y_{z}$ contains the content of $z$ with $Y$ style. We use U-net~\cite{ronneberger2015u} as the architecture of all the generators and PixelGAN~\cite{isola2017image} for the discriminators.

\subsection{Cycle-adversarial Loss}
 The basic adversarial loss is expressed as:

\begin{equation}
\begin{split}
\mathcal{L}_{GAN}(G, D_{Y}, X, Y)= & \mathbb{E}_{y \sim p_{data}(y)}[logD_{Y}(y)]\\ + & \mathbb{E}_{x \sim p_{data}(x)}[1-D_{Y}(G(x))]
\end{split}
\label{equ:equ1}
\end{equation}
After defining the loss function, we optimize it as \[min_{G}max_{D_{Y}}\mathcal{L}_{GAN}(G, D_{Y}, X, Y),\] which means that we maximize the distance between the generated images and HDTV frames by training $D_{Y}$ and then minimize the distance between them by training $G$ to generate images look similar to HDTV frames. Using the same training strategy, we optimize the other adversarial loss, \[min_{F}max_{D_{X}}\mathcal{L}_{GAN}(F, D_{X}, Y, X).\]

Owing to lacking ground truth, it is no guarantee that the content of the input is unchanged after translating. If we train the model long enough with adversarial loss, the generated image would become an image that is independent of input. Hence, a cycle consistency loss is essential to keep the information of input. The cycle consistency loss can be formulated as
\begin{equation}
\begin{split}
\mathcal{L}_{cyc}(G, F)=& \mathbb{E}_{x \sim p_{data}(x)}[\lVert F(G(x))-x\lVert_{1} ] \\ + & \mathbb{E}_{y \sim p_{data}(y)}[\lVert G(F(y))-y\lVert_{1}], \\ &
\end{split}
\label{equ:equ2}
\end{equation} 
where $\, ||*||_{1}  \,$ denotes $\, L_1\,$ norm.

\subsection{Resolution Loss}
 CycleGAN~\cite{zhu2017unpaired} does not consider the change of resolution during the translation but in our problem, the change of resolution cannot be ignored. However, VHS frames have no corresponding high-resolution or clear images as ground truth. Thus we synthesize data $Z$ by blurring and downscaling HDTV frames with the purpose of making them close to the resolution of VHS frames. In order to fuse data from $X$ and $Z$, we train them with the same generator $G$ by sharing their weights. In addition, we use perceptual loss to constrain the perceptual quality between output and target by reducing their distances of deep features extracted by a pre-trained VGG19~\cite{simonyan2014very} network. The perceptual loss can be expressed as:
\begin{equation}
\begin{split}
\mathcal{L}_{perc}(G)=\mathbb{E}_{y \sim p_{data}(y)}[\lVert VGG(G(z))- VGG(y) \lVert_{2}]
\end{split}
\label{equ:equ3}
\end{equation}
where $\, ||*||_{2}  \,$ denotes $\, L_2 $ norm.

Similar to the adversarial loss in Eq.~\ref{equ:equ1}, we define another adversarial loss, which distinguishes the generated images and HDTV frames in resolution. As input is blurred HDTV frames, the difference between output and target is only the resolution that $D_{Z}$ need to capture. Our goal is just like above:
\[min_{G}max_{D_{Z}}\mathcal{L}_{GAN}(G, D_{Z}, Z, Y),\] which makes translated images look more like high-resolution ones.

\subsection{Overall Objective}
The overall objective function of our multi-task learning model is
\begin{equation}
\begin{split}
\mathcal{L}(G,F,D_{X},&D_{Y}, D_{Z})= \mathcal{L}_{GAN}(G,D_{Y},X,Y)  \\+& \mathcal{L}_{GAN}(F,D_{X},Y,X) +  \lambda\mathcal{L}_{cyc}(G,F) \\+& \mathcal{L}_{GAN}(G, D_{Z},Z,Y)  + \kappa\mathcal{L}_{perc}(G)
\end{split}
\label{equ:equ4}
\end{equation}
where $\mathcal{L}_{GAN}(G,D_{Y},X,Y)$ , $\mathcal{L}_{GAN}(F,D_{X},Y,X)$ and $\mathcal{L}_{cyc}(G,F)$ are the losses that constrain the image style while the remaining losses constrain the image resolution. We aim at optimizing the following:
\begin{equation}
\{ G^{*}, F^{*} \} = arg\min \limits_{G,F}\max \limits_{D_{X},D_{Y},D_{Z}}\mathcal{L}(G,F,D_{X},D_{Y},D_{Z})
\label{equ:equ5}
\end{equation}

\section{Results and Evaluation}
\label{sec:sec4}

\subsection{Training Details}
We collected a dataset from the Internet. The VHS frames were taken from a 1990s program of the show "Late Show with David Letterman" and the HDTV frames were taken from a 2010s program of the same show. We also collected VHS frames from the 1995 production of 'Toy Story' and HDTV frames of the 2010 production of the same title. There are around 8000 VHS frames and around 8000 HDTV frames, $95\%$ were used as training data and the rest as testing data.

Since the improvement of resolution is more difficult than that of colors, for each training iteration of the cycle-adversarial loss, we train the resolution loss several iterations (5 in the experiment). Our learning rate is 0.0001 and we use Adam to optimize the model while momentums are set as 0.5 and 0.999. And the parameters in Eq.~\ref{equ:equ4} is set as $\lambda$ = 0.1 and $\kappa$ = 0.05.

\begin{figure*}[!ht]

\begin{minipage}[b]{1.0\linewidth}
  \centering
  \centerline{\includegraphics[width=12cm]{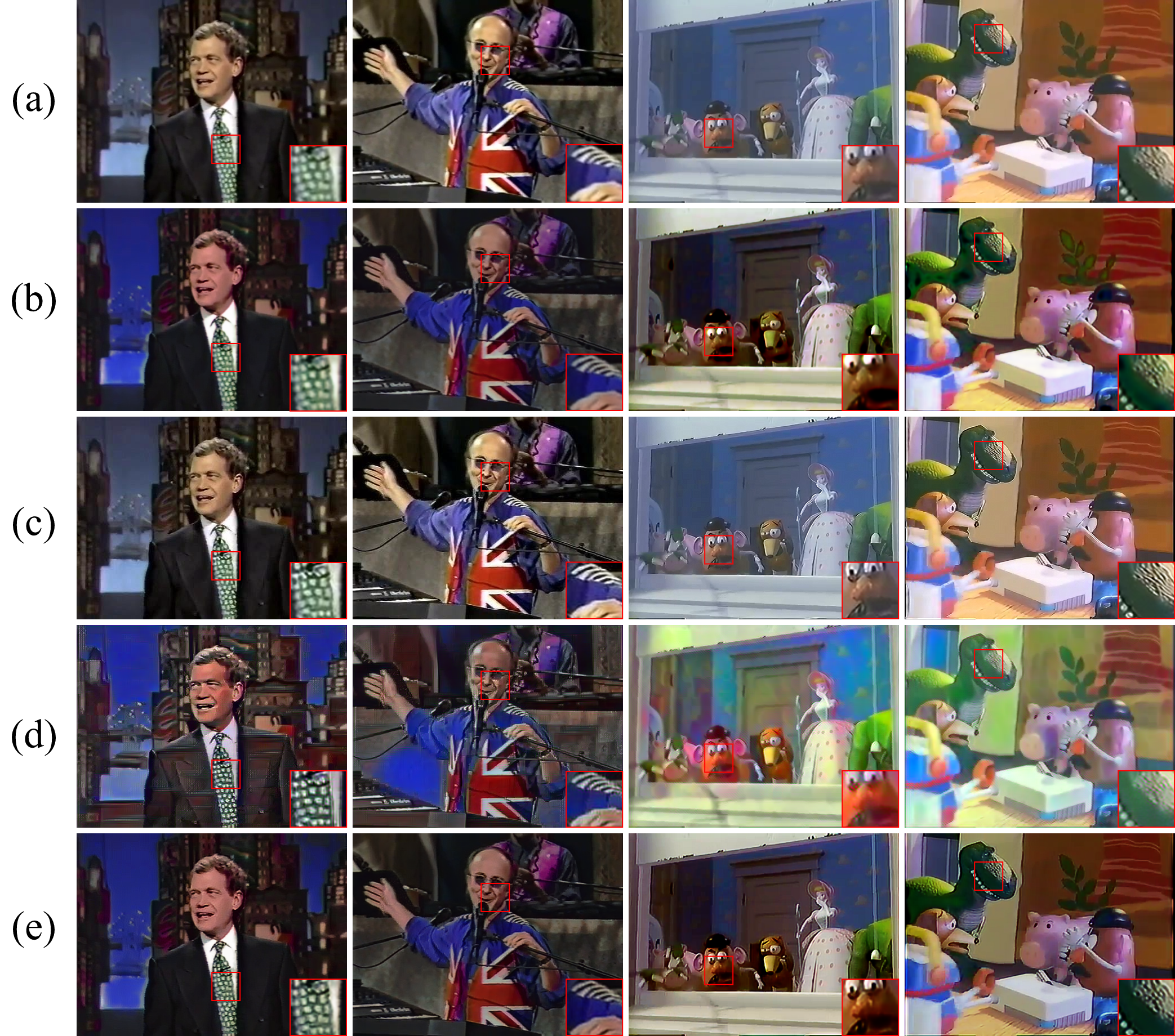}}
  
\end{minipage}
\caption{Visual examples of translating  input (VHS) to HDTV by various methods. These five rows represent (a)input, (b)CycleGAN~\cite{zhu2017unpaired}, (c)DRRN~\cite{tai2017image}, (d)Johnson et al.~\cite{johnson2016perceptual}, (e)ours respectively.}
\label{fig:res4}
\end{figure*}

\subsection{Visual Results}
To show the superiority of the proposed model, we compare it with CycleGAN~\cite{zhu2017unpaired}, DRRN~\cite{tai2017image}, and Johnson et al.~\cite{johnson2016perceptual}. In Fig.~\ref{fig:res4}, the results demonstrate that CycleGAN~\cite{zhu2017unpaired} and Johnson's~\cite{johnson2016perceptual} method can improve color contrast, but CycleGAN~\cite{zhu2017unpaired} fails to improve the resolution of images and Johnson's~\cite{johnson2016perceptual} method causes some artifacts and the color seems strange. On the other hand, DRRN~\cite{tai2017image} enhances image resolution but the color remains unchanged. Our method performs better both in color and resolution.

\subsection{Quantitative Evaluation}
Because our problem is lack of ground truth, we can only use some No-Reference Image Quality Assessment(NRIQA) methods to evaluate our results quantitatively. There are two common methods called BRISQUE~\cite{mittal2012no}(blind/referenceless image spatial quality evaluator) and PIQUE~\cite{venkatanath2015blind}(perception-based image quality evaluator). 
Comparison of various methods is shown in Table~\ref{tab1}.
The lower the score, the better the quality.

\begin{table}
\caption{Comparison of several methods in BRISQUE and PIQUE score.}\label{tab1}
\setlength{\tabcolsep}{2.5mm}{
\begin{tabular}{|l|l|l|l|l|l|}
\hline
evaluation methods & input & CycleGAN & DRRN & Johnson et al. & ours\\
\hline
BRISQUE & 31.4105 & 30.43 & 29.34 & 26.03 & {\bfseries 24.70}\\
PIQUE & 62.1714 & 59.91 & 52.96 & 51.29 & {\bfseries 48.55}\\
\hline
\end{tabular}}
\end{table}

In addition, subjective evaluation is another important evaluation for no-reference image quality assessment. In the subjective evaluation procedure, we show 5 images, input and four generated images by four methods, to 50 participants, and let them select their preferred image under unified environment. Collect their selection and compute the preference percentage of 5 kinds of images. Table~\ref{tab2} shows the voting results. We can see that there is an obvious preference for our method against all other methods for the translating task.

\begin{table}
\caption{Comparison of several methods in human preferred score.}\label{tab2}
\setlength{\tabcolsep}{2.9mm}{
\begin{tabular}{|l|l|l|l|l|l|}
\hline
 & input &  CycleGAN & DRRN & Johnson et al. & ours\\
\hline
human preferred & 0\% & 7.94\% & 9.99\% & 13.52\% & {\bfseries 68.55\%}\\

\hline
\end{tabular}}
\end{table}

\section{Concluding Remarks}
\label{sec:sec5}
In this paper, we focus on an unsupervised multi-task problem, VHS video to HDTV video translation, which takes color contrast as well as spatial resolution into account. In order to incorperate multiple tasks into a single model and handle the unsupervised problem, we propose a multi-task adversarial learning model, which learns not only on color contrast but also image resolution. Taking advantage of cycle-adversarial loss and resolution loss, we fuse two kinds of input images by sharing weights between two networks. Experimental results demonstrate the effectiveness of our method.  

\section{Future Works}
\label{sec:sec6}
So far, our work focus on video translation frame by frame. This can be limited in that neighboring frames might not be consistent in terms of colors and contrast after translation, which would lead to a not very enjoyable video. For this reason, we will take into account the continuity of video, just like how Wang et al.~\cite{wang2018video} have done on video-to-video synthesis.

\subsection*{Acknowledgement}
This work was supported by initial funding of newly-introduced teacher in Shenzhen University with No. 2019121. The authors would like to thank the editors and reviewers for their constructive suggestions on our work. The corresponding author of this paper is Fei Zhou.
\bibliographystyle{splncs04}
\bibliography{MMM2020}

\end{document}